\documentclass{emulateapj}
\usepackage[latin1]{inputenc}
\usepackage{natbib}
\defcitealias{richards:belJ1004}{R04}
\defcitealias{inada:j1004}{I03}

\slugcomment{To appear in ApJL}

\shorttitle{INTEGRAL Field Spectroscopy of SDSS 1004+4112}
\shortauthors{Gómez-Álvarez et al.}

\begin{document}


\title{Recurrence of the Blue Wing Enhancements in the High
Ionization Lines of SDSS 1004+4112 A}


\author{P. Gómez-Álvarez\altaffilmark{1}, E. Mediavilla\altaffilmark{1}, 
J. A. Muñoz\altaffilmark{2}, S. Arribas\altaffilmark{3,8,1,7},
S.F. Sánchez\altaffilmark{4,5,6}, A. Oscoz\altaffilmark{1}, 
F. Prada\altaffilmark{4}, M. Serra-Ricart\altaffilmark{1}}






\altaffiltext{1}{Instituto de Astrof\'{\i}sica de Canarias, V\'{\i}a L\'actea S/N,  E-38200 La Laguna, Tenerife, Spain}
\altaffiltext{2}{Departamento de Astronom\'{\i}a y Astrof\'{\i}sica, Universidad de Valencia,  E-46100 Burjassot, Valencia, Spain}
\altaffiltext{3}{Space Telescope Science Institute, 3700 San Martin Drive, Baltimore, MD 21218, USA}
\altaffiltext{4}{Instituto de Astrof\'{\i}sica de Andalucia (CSIC), E-18008 Granada, Spain}
\altaffiltext{5}{Astrophysikalisches Institut Postdam, And der Sternwarte 16,
14482 Postam, Germany}
\altaffiltext{6}{Calar Alto Observatory, CAHA, Apdo. 511, E-04044 Almería, Spain}
\altaffiltext{7}{Consejo Superior de Investigaciones Cient\'{\i}ficas (CSIC), Spain}
\altaffiltext{8}{Affiliated with the Space Telescope Division of the European Space Agency, ESTEC, Nordwijk, Holland}



\begin{abstract} 

We present integral field spectroscopic observations of the quadruple-lensed QSO 
SDSS 1004+4112 taken with the fiber system INTEGRAL at the William Herschel 
Telescope  on   2004 January 19.
In May 2003 a blueward enhancement in the high ionization lines of SDSS 1004+4112A 
was detected and then faded. Our observations are the first to note a 
second event of similar 
characteristics less than one year after. Although initially attributed to 
microlensing, the resemblance among the spectra of both events and the absence 
of microlensing-induced changes in the continuum of component A are puzzling. 
The lack of a convincing explanation under the microlensing or intrinsic
variability hypotheses makes the observed enhancements particularly relevant,
calling for close monitoring of this object.

%
%
%


%

\end{abstract}




\section{Introduction}
The large separation (14.62 arcsec) quadruple gravitationally lensed quasar was first identified
by \citet{inada:j1004} (hereinafter \citetalias{inada:j1004}) in the Sloan Digital Sky Survey. The derived redshift is
 $z=1.732$ and the lensing complex consists of a cluster of galaxies at $z=0.68$.

The components of a multiple lensed system can show uncorrelated variability due to the
gravitational (de)magnification produced by compact deflectors aligned with the source. This phenomenon,
microlensing \citep{chang1979,chang1984}, is routinely observed in the continuum emission of multiple imaged
QSOs, and it is considered a powerful tool for mapping 
the QSO continuum source \citep{yonehara:1999}.

According to our current understanding of the structure of active galactic 
nuclei (AGNs), the continuum source of a QSO is surrounded by a wider region, 
where the broad emission lines (BELs)  originate. 
Early versions of the standard model of AGNs assumed a too large size for 
the broad line region (BLR) ($\sim$1 pc) to be microlensed by solar-mass objects 
\citep{nemiroff:agnBLR,Schneider:qsoBLR}.  However, these early estimates 
were reduced by two orders of magnitude by the new measurements of the BLR size
 based in the reverberation method \citep{wandel:rever,kaspi:rever}. 
 Considering the new size estimates, a variety of theoretical models for the kinematic 
 structure of the BLR \citep{popovic:microlensing,cristina:blr,lewis:microlensing}
 have shown that microlensing by stellar sized objects can produce significant 
 amplifications of the BEL of multiple imaged QSOs, especially in the high
 ionization lines.

\citet{richards:belJ1004} (hereinafter \citetalias{richards:belJ1004})
 found excess in the blue wings of several high ionization lines of SDSS
 1004+4112A, the highest magnified lensed image, relative to the same lines in the other
 images. The excess persisted for at least 28 days and then faded. No such
 enhancements in the blue wings were seen in the low ionization lines. 
These results are in agreement with current ideas about the stratification 
of the BLR according to the ionization level \citep{peterson:supermassiveBH}
 and suggest the possibility of scanning the BLR with high spatial resolution. 
There is no evidence of continuum microlensing in the SDSS 1004+4112 event 
(however see \citealt{ota:chandra}). This lack of correlation can be explained 
for a particular event \citep{lewis:microlensing}, but the apparent absence 
of continuum amplification in all the reported cases of BEL microlensing 
(\citetalias{richards:belJ1004},\citealt{chartas:J0414,chartas:H1413,dai:qso2237})
 is intriguing. 

In this article we present new spectroscopic measurements of SDSS 1004+4112 that 
show a recurrence of the blueward magnification of the high ionization line wings 
of image A.






\section{Observations and data reduction}
SDSS 1004+4112 was observed on  2004 January 19 with the fiber system 
INTEGRAL \citep{arribas:integral}, at the 4.2 m William Herschel Telescope (WHT, Roque de
los Muchachos Observatory, La Palma, Spain). 
Two sets of three 1800 s exposures each were taken corresponding to 
two pointings separated by few arc seconds. We use the standard fiber bundle \#3 
(SB3) which consists of 115 object + 20 sky fibers, each one subtending 2.70$''$ in diameter
and covering a sky area of $33.6'' \times 29.4''$. A 300$\rm mm^{-1}$ 
diffraction grating was used, giving a spectral coverage of $\sim$6000 \AA\ 
centered at 5700 \AA\ and a spectral resolution of 19 \AA, 
with a spectral dispersion of 6.2 \AA\ $\rm pix^{-1}$. The observations were 
performed under medium seeing (1.5$''$--2$''$) conditions. The data 
reduction procedures included cosmic ray rejection, aperture tracing, 
spectra extraction, wavelength calibration, throughput correction, 
sky subtraction, atmospheric extinction correction and flux calibration. Since the 
vast majority of the flux is concentrated in one or two fibers, the  spectrum 
has been obtained directly by combining their fluxes.

The absolute flux calibration must be treated with caution since the calibration 
star position within the fiber face could affect the amount of flux lost in 
the interfiber regions. 
The goodness of our relative photometry is shown by the comparison with the photometry of 
I03 (Figures \ref{ABMagnitudes} and \ref{ACMagnitudes}).

Apart from WHT observation we have use data from the Keck and ARC telescopes (See Table
\ref{Observations}), which has been kindly provided by Gordon T. Richards.

\begin{deluxetable}{llllll} 
\tabletypesize{\scriptsize}
\tablecaption{Summary of SDSS 1004+4112 observations}
\tablehead{
\colhead{Date}& 
\colhead{$\Delta T$\tablenotemark{a}} & 
\colhead{Telescope} & 
\colhead{Ref.} & 
\colhead{Components}
}
\startdata
2003 May 31 & 	0 		& Keck	&   \citetalias{richards:belJ1004} &	ABCD\\
2003 Nov 21 & 	174		& ARC\tablenotemark{c}		&		\citetalias{richards:belJ1004} &	AB\\
2003 Nov 30 & 	183		& ARC		&		\citetalias{richards:belJ1004} &	AB\\
2003 Dic 1	& 	184		& ARC		&		\citetalias{richards:belJ1004} &	AB\\
2003 Dec 22 & 	205 	&	ARC		&		\citetalias{richards:belJ1004} &	AB\\
2004 Jan 19 & 	233		& WHT		&	  -															 & ABCD\\
2004 Mar 26 & 	300		& ARC		&		R2\tablenotemark{b} 					 & AB\\ 
\enddata
\tablenotetext{a}{Elapsed days from first observation.}
\tablenotetext{b}{See \cite{richards:mlevent}}
\tablenotetext{c}{Astrophysical Research Consortium (ARC) Telescope
}
\label{Observations}
\end{deluxetable}
\begin{figure}[ht]
\includegraphics[width=8.5cm]{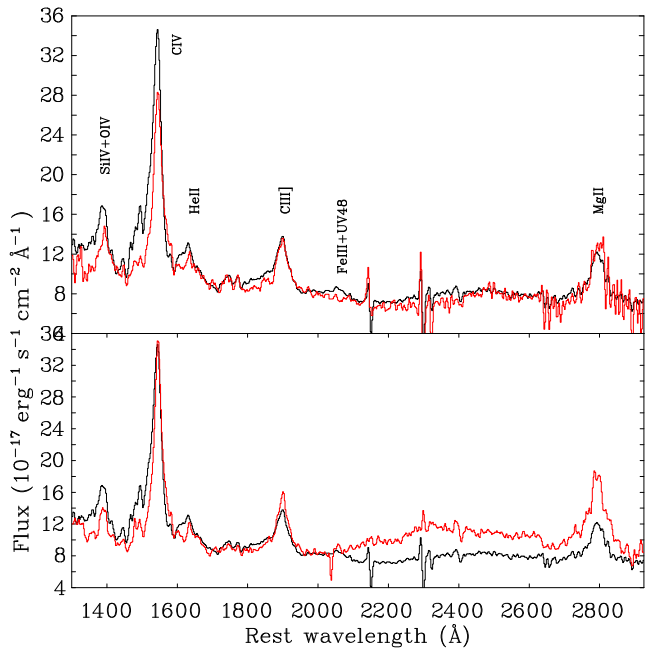}
\caption{Comparison between spectra of components A (black in both panels), 
B (upper panel), and  C (lower panel) for SDSS 1004+4112 at the WHT. Spectra B 
and C have been scaled to normalize its continuum to that of A in the 
1700--1800 \AA\ wavelength range.}
\label{ABSpectra}
\end{figure}


\section{Results}
\subsection{Comparison of A, B, and C spectra}
Inspection of Figure \ref{ABSpectra} reveals that: i) component A shows 
enhancements relative to components B and C in the blue wings of the high 
ionization emission lines (Si IV/O IV], C IV, He II, Al III, and UV48 Fe III 
complex) similar to those found by \citetalias{richards:belJ1004};  ii) the
 A and B continua match very well in all the observed range although the C 
 continuum exhibits an excess beyond 2000\AA; iii) the low ionization lines, 
 C III] and Mg II, match reasonably well with the same normalization factor 
 as the continua; and iv) the EWs of the low ionization lines 
C III] and Mg II are larger in C than in A, and the EW of the C IV line 
is larger in C than in B.

\subsection{Comparison with spectra of other epochs}
\begin{figure}[h]
\includegraphics[trim=50 90 0 0, angle=-90, width=8.5cm]{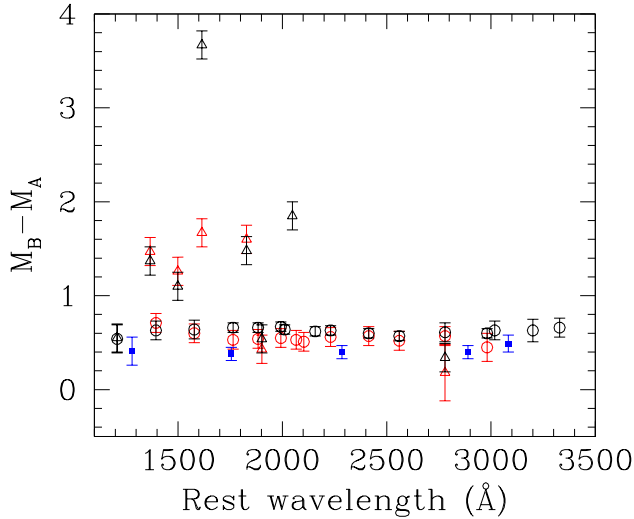}
\caption{$B-A$ magnitude difference curve. Triangles: emission 
lines. Circles: continua. Blue squares: photometric data from \citet{inada:j1004}. Red: 
WHT-INTEGRAL. Black: Keck-LRIS.}
\label{ABMagnitudes}
\end{figure}

\begin{figure}[h]
\includegraphics[trim=50 90 0 0, angle=-90, width=8.5cm]{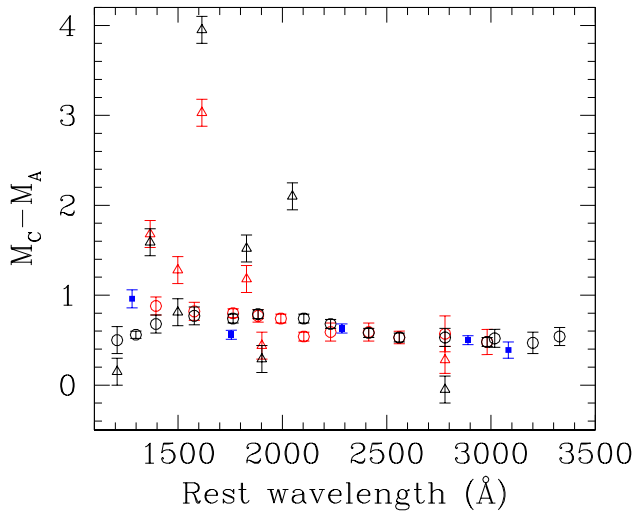}
\caption{$C-A$ magnitude difference curve. Triangles: emission 
lines. Circles: continua. Blue squares: photometric data from \citet{inada:j1004}. Red: 
WHT-INTEGRAL. Black: Keck-LRIS.}
\label{ACMagnitudes}
\end{figure}

Seven spectroscopic observations were made on between 2003, May 31 and 2004,
March 26 (See Table \ref{Observations}). 
The differences between the WHT spectra corresponding to components A and B 
are remarkably similar to those reported by \citetalias{richards:belJ1004}
 and \citet{oguri:j1004} from Keck observations taken on  2003 May 31. 
A direct comparison, not  shown here (See figures \ref{ABMagnitudes} 
and \ref{ACMagnitudes}), among our spectra and the Keck 2003 data shows a very good global 
agreement for components A, B, and C. 
In the case of component A, some differences are found 
in C IV  (the EW is larger in the WHT spectrum) and in He II  
(the blueward asymmetry is stronger in the Keck spectrum).  

\begin{figure}[h]
\includegraphics[width=8.5cm]{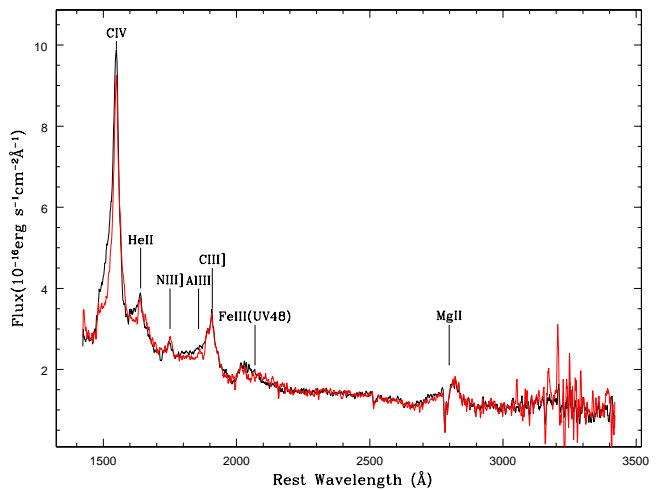}
\caption{SDSS 1004+4112 A/B spectra taken on 26 March 2004 at the ARC
telescope.}
\label{26MarSpectrum}
\end{figure}

A direct comparison  (not shown here) 
of the A and B ARC spectra based on the normalization of the continua shows that 
during the epoch  2003 November 21 to  
 2003 December 22 the blue excess disappears in the line of highest ionization (He II) 
and severely declines in CI V but it is still present in the lower 
ionization Al III and Fe III UV48 lines. 
The WHT observations show that the blue enhancements in  component A 
again reached (on  2004 January 19)  a level comparable to that of the Keck 
observations (2003 May 31). In Figure \ref{26MarSpectrum} we see that on 
 2004 March 26 the system shows similar features.

\subsection{B-A and C-A magnitude difference curves}
We estimate the flux ratio independently for the continuum
and the  emission lines. For the high ionization lines we have integrated 
only the blue half-side to enhance the asymmetries. 
To obtain the emission line flux we subtract a linear
interpolation of the continuum from the regions adjacent to the lines.  
We then computed the continuum flux over intervals of 200 \AA.
The resulting magnitude differences, 
$\Delta m_{BA}=m_B-m_A$ and $\Delta m_{CA}=m_C-m_A$, for both the continuum and the 
emission lines are presented in Figures \ref{ABMagnitudes} and \ref{ACMagnitudes}. 

In Figure \ref{ABMagnitudes} we present the WHT and Keck data. The agreement between 
both observations is quite remarkable. The small offset ($\sim$0.05 mag) 
between the Keck and WHT continuum difference curves strongly constrains 
microlensing-induced continuum variability between 
both epochs (2003 May 31 and 2004 January 19). There is also  good 
agreement (better than 0.05 mag with respect to the WHT continuum curve) with 
the photometric data by I03. This is reasonable also in terms 
of intrinsic variability if we take into account that the yearly intrinsic 
variation of this object could be of about 0.2 mag (according to the preliminary
results of the photometric monitoring our group is performing), and that the 
predicted time delay between components A and B is about 1 month. 

The $B-A$ differences for the low ionization lines (C III], Mg II) 
match (within errors) the continuum difference curve  (there is some marginal 
evidence of a negative offset for the Mg II). This result does not 
support the existence of microlensing in the continuum. The blue half-sides of 
the high ionization lines are above the continuum with an offset of about 1 
mag for the Si IV, C IV, Al III, and the UV48 Fe III complex. The blue 
half-side of the He II line shows a larger offset of about 3 mag respect 
to the continuum. Finally, there is no evidence of wavelength dependence 
of the continuum differences curve that have been invoked in several cases 
\citep[see][]{Wucknitz::1993, motta:Q0909} to justify chromatic 
microlensing. This also excludes the possibility of differential dust extinction.

The $C-A$ difference magnitude curves are shown in Figure \ref{ACMagnitudes}. 
The agreement between the WHT difference curve for the continuum and the 
photometric data by I03 is also remarkably good and does not support the 
existence of noticeable microlensing variability during the period between 
both observations. The Keck spectrum needed an offset of 0.8 mag to be matched
to that of the I03 data. 
In Figure \ref{ACMagnitudes} the high ionization 
lines appear above the continuum but the low ionization lines do not 
match the continuum curve, which is $\sim 0.45$ mag above. An offset between 
the line and continuum quotients of two components of a gravitational lens has 
been interpreted as  evidence for microlensing 
\citep[see][]{Evencio:0909,motta:Q0909}.
If we assume that this hypothesis can be applied to the low ionization BEL and take 
into account that microlensing in the component A is for the same reason 
unlikely, we should accept that the continuum of component 
  C is demagnified by microlensing. 
	On the other hand, the 
  continuum difference curve is not flat but shows a relative decrease in C 
  with respect to A for shorter wavelengths (the color difference between components A and C 
   has already been reported during the identification of the system as a 
  multiple imaged QSO of large separation, see I03). This can be 
  easily interpreted as dust extinction of component C (see e.g. Falco et~al.
1999 \& Mu\~noz et~al.  2004). 
An interesting alternative possibility to explain the color differences around
2100\AA\ and longer (See Figure \ref{ABSpectra}) is UV FeII emission microlensed (magnified) in the C
image and not in the others. The empirical iron emission template given by
Vestergaard \& Wilkes (2001) matches qualitatively the observed variation 
with wavelength.
Both possibilities can be discriminated with future observations, since
microlensing in the iron lines would show time variability but extinction
would not.


\section{B-A variability for the HeII and CIII] emission lines}
\begin{figure}[ht]
\centering
\includegraphics[angle=270,width=8.5cm]{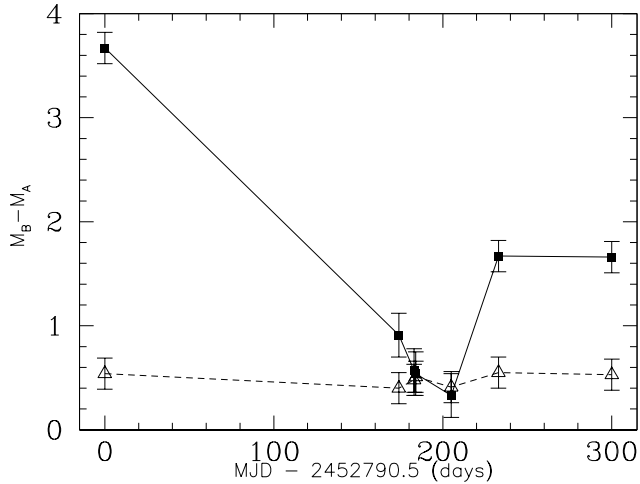}
\caption{Magnitude differences between component A and B for CIII] (triangles) 
and He II lines (squares)}
\label{BAVariability}
\end{figure}

In Figure \ref{BAVariability} we present the $B-A$ values for the C III] and the HeII (blue
half-side) emission lines corresponding to the seven epochs with spectra
available (see Section 3.2). The $B-A$ value for the low ionization  C III]
line is almost constant (matching the $B-A$ value for the continuum; see
previous section), whereas the He II emission line  exhibits strong variability 
from a maximum  during the first epoch, a posterior decay, and a final rise. 
It is very noticeable that the minimum of $B-A$ for the He II coincides with 
the baseline defined by the CIII].

\section{Discussion and conclusions}
The interpretation of the apparent recurrence of the blueward enhancement of the
high ionization lines is challenging. If the enhancements were produced by
intrinsic variability, they should appear with some delay in all components. However, observations do not reveal changes in the spectrum B that
could be related to the blue wing asymmetries detected in A. On the other hand,
the lack of continuum variability that may induce the change in the emission
lines is also against this hypothesis.

Alternatively, the rising and fading of the blue wing enhancements can be
explained by the relative displacement of a caustic network with respect to the
BLR. However the displacement of the caustic across 
a kinematically structured BLR should induce velocity changes in the high
ionization lines \citepalias{richards:belJ1004} rather than recurrences like
the one observed. Another inconvenience of the microlensing hypothesis is the
very short time-scale for emission line variability. According to the
observations, the He II blue wing excess was not present in the ARC  2003
December 22 spectrum but appears in the WHT spectrum taken on 2004 January 19
(See Figure \ref{BAVariability}) implying a rise time of less than one month,
which reduces by a factor of $\sim$6 the upper limit inferred by
\citetalias{richards:belJ1004} for the size of the amplified region under the
hypothesis of microlensing. Richards et al.\ (2004b) estimated that the
impact of microlensing in the A continuum was no more than 20\% during the time
spanned by their observations. Our comparison (see previous section) among the
\citet{inada:j1004}, Keck, and WHT data indicates no traces of $B-A$ continuum
 variability during the period between 2003 May 2 and 2004 January 19. 
On one hand, this implies that the continuum microlensing time scale (if any),
 should be much larger than the one associated with high ionization line 
 variability. On the other hand, (a) the agreement among the $B-A$ data corresponding to the low 
ionization lines and the continuum and (b) the apparent absence of microlensing
 chromaticity do not support this hypothesis. 
Thus, it seems that the process causing the variability of the high ionization 
lines does not affect the continuum. According to 
\citet{lewis:microlensing} and \citetalias{richards:belJ1004}, microlensing by a
caustic network can cause uncorrelated variability between the high ionization
lines and the continuum. However, it is difficult to explain the absence of
significant continuum microlensing ($\Delta (B-A) < 0.05$ mag) when the high
ionization lines have undergone a complete cycle that in the case of the blue
wing of He II implies $\Delta (B-A)$ differences of several magnitudes.

The decoupling between continuum and emission lines can be explained by
supposing that the emission line region is formed by clouds of ionized gas
moving at high velocity and/or considering a special geometry (biconical) for 
the BLR. This would favor the appearance of asymmetrical features and might
explain the uncorrelated variability between emission lines and continuum. Even
the possibility of microlensed transient events as SNs or shocks deserves 
consideration. We will explore these possibilities in a future work.
\acknowledgments
We are grateful to Gordon T. Richards for providing previous observations and
 useful comments.
We also wish to acknowledge the referee, Ian Browne for valuable comments and 
ideas. 
This work was supported by the European Community's Sixth
Framework Marie Curie Research Training Network Programme,
Contract No.  MRTN-CT-2004-505183 ``ANGLES'', and by the Ministerio de
Educaci\'on y Ciencia of Spain with the grants AYA2004-08243-C03-01 and
AYA2004-08243-C03-03.

\bibliographystyle{apj}

\clearpage

\clearpage

\clearpage


\end{document}